\documentclass[letterpaper,twocolumn,10pt]{article}
\usepackage{usenix2019_v3}

\usepackage{amsmath,amssymb,amsfonts}
\usepackage{algorithmic}
\usepackage{graphicx}
\usepackage{textcomp}
\usepackage[table,xcdraw]{xcolor}
\usepackage[final]{hyperref}
\usepackage{subcaption}
\usepackage{makecell}
\usepackage[font=small,format=plain,labelfont=bf]{caption}
\usepackage{balance}
\usepackage{xspace}
\usepackage{fancyref}
\usepackage{dblfloatfix}
\usepackage{tabularx}
\usepackage[compact]{titlesec}
	\titlespacing{\section}{0pt}{2ex}{1ex}
	\titlespacing{\subsection}{0pt}{1ex}{1ex}
	\titlespacing{\subsubsection}{0pt}{1ex}{0ex}
	\titlespacing{\paragraph}{0pt}{1.25ex plus 1ex minus .2ex}{1em}
\usepackage{enumitem}
\usepackage{titling}
\usepackage{tikz}
\usepackage{microtype}
\usetikzlibrary{arrows}

\graphicspath{{./figures/}}

\titleformat*{\section}{\large\bfseries}
\titleformat*{\subsection}{\normalsize\bfseries}
\titleformat*{\subsubsection}{\normalsize\bfseries}
\titleformat*{\paragraph}{\normalsize\bfseries}
\titleformat*{\subparagraph}{\normalsize\bfseries}

\AtBeginDocument{%
	\abovedisplayshortskip=0pt plus 3pt
}

\posttitle{\par\end{center}}

\setlist[itemize]{noitemsep, topsep=0pt}
\setlist[enumerate]{noitemsep, topsep=0pt}

\definecolor{lightgray}{rgb}{0.88,0.88,0.88}
\definecolor{darkred}{rgb}{0.57,0.13,0.28}

\hypersetup{
	colorlinks   = true, 
	urlcolor     = darkred, 
	urlbordercolor=white, 
	linkcolor    = darkred, 
	linkbordercolor = white, 
	citecolor    = darkred, 
	citebordercolor = white, 
	breaklinks   = true, 
}
\makeatletter
\Hy@AtBeginDocument{
	\def\@pdfborder{0 0 1} 
	\def\@pdfborderstyle{/S/U/W 0.5} 
}
\makeatother

\pgfsetarrows{latex-latex}
\tikzset{%
	base/.style = {inner sep=5pt,
		text centered,
		thin,
		font=\rmfamily},
	round/.style = {base,
		rectangle,
		rounded corners=1ex,
		draw=black,
		fill=gray!20,
		minimum height=0.35in},
	opportunity/.style = {round,
		fill=gray!5,
		draw=gray!90,
		text=black!60}
}

\usepackage[
style = trad-abbrv,
backend      = biber,
citestyle    = numeric-comp, 
mincrossrefs = 1000, 
]{biblatex} 
\bibliography{proceedings.bib,paper.bib}

\DeclareSourcemap{
	\maps[datatype=bibtex, overwrite]{
		\map{
			\step[fieldset=editor, null]
			\step[fieldset=location, null]
			\step[fieldset=publisher, null]
			\step[fieldset=isbn, null]
			\step[fieldset=issn, null]
			\step[fieldset=pages, null]
		}
	}
}

\date{}

\title{\Large \bf Identifying Disinformation Websites Using Infrastructure Features}

\author{
{\rm Austin Hounsel}\\ Princeton University
\and
{\rm Jordan Holland}\\ Princeton University
\and
{\rm Ben Kaiser}\\ Princeton University
\and
{\rm Kevin Borgolte}\\ Princeton University
\and
{\rm Nick Feamster}\\ University of Chicago
\and
{\rm Jonathan Mayer}\\ Princeton University
}

\begin{document}

\maketitle

\thispagestyle{empty}


\begin{abstract}
Platforms have struggled to keep pace with the spread of disinformation. 
Current responses like user reports, manual
analysis, and third-party fact checking are slow and difficult to scale,
and as a result, disinformation can spread unchecked for some time after
being created. Automation is essential for enabling platforms to respond
rapidly to disinformation.

In this work, we explore a new direction for
automated detection of disinformation websites: infrastructure features.
Our hypothesis is that while disinformation websites may be perceptually
similar to authentic news websites, there may also be significant
non-perceptual differences in the domain registrations, TLS/SSL
certificates, and web hosting configurations. Infrastructure features are
particularly valuable for detecting disinformation websites because they
are available before content goes live and reaches readers, enabling early
detection.

We demonstrate the feasibility of our approach on a large corpus of labeled
website snapshots. We also present results from a preliminary real-time
deployment, successfully discovering disinformation websites while
highlighting unexplored challenges for automated disinformation detection.
\end{abstract}

\section{Introduction} \label{sec:introduction} In recent years, the Internet
has made disinformation cheaper, easier, and more effective than ever
before~\cite{arXiv:1804.08559}. The same technologies that have democratized
online content creation, distribution, and targeting are increasingly being
weaponized to mislead and deceive. Russia deployed disinformation
to interfere in the 2016 U.S. presidential election~\cite{ssci2019report2,
mueller2019report1, doj2018ira-indictment, diresta2018tactics, howard2018ira,
zannettou2019disinformation-warfare}, and political disinformation campaigns
have also struck dozens of other nations~\cite{bradshaw2019global,
bradshaw2018challenging}. Other disinformation campaigns have been
economically motivated, driving page views for advertising revenue, pushing
products, or undermining competitors~\cite{marwick2017media}.

The major online platforms have not kept pace with the spread of
disinformation. Responses to disinformation mostly rely on user reports,
manual analysis, and fact checking, which are slow and difficult to
scale~\cite{ananny2018partnership}. Similarly, previous work on automated 
detection of disinformation has focused on textual and social graph features, 
which often rely on disinformation already being shared 
~\cite{potthast2018stylometric,afroz2012detecting,conroy2016automatic,kumar2016disinformation,gupta2013faking,zhao2015enquiring,yang2012automatic,ma2017detect,wu2015false}.
These responses give disinformation an
asymmetric advantage, enabling it to spread and affect perceptions in the
hours and days after it is first distributed---a critical period during which
disinformation may be most effective~\cite{zubiaga2016analysing}.

In this paper, we explore the use of infrastructure features to detect
disinformation websites. Our hypothesis is that while disinformation websites
may be perceptually similar to authentic news websites, there may be
significant non-perceptual differences in the domain registrations, TLS/SSL
certificates, and web hosting configurations.

We are motivated by prior work
in the information security field that demonstrated viable early detection for
malware, phishing, and scams using machine learning and a combination of
carefully engineered network-level and application-level
features~\cite{hao2016predator, wang13detecting, thomas2011design,
hao2009detecting}. We use similar insights to support discovery of
disinformation websites based on infrastructure features. These
features are particularly valuable because they are available before
disinformation campaigns begin.

We construct features derived from a website's domain, certificate, and hosting
characteristics and then apply multi-label classification to categorize the website as
disinformation, authentic news, or other (i.e., lacking news content). Our
evaluation shows the feasibility of our approach on a large, labeled dataset of
website snapshots. We also present results from a preliminary real-time
deployment, in which we were able to discover previously unreported disinformation websites. We outline the challenges for automated disinformation detection based on our
experience.

\begin{figure*}[t!] \centering \resizebox{\textwidth}{!}{
  \renewcommand{\baselinestretch}{1.5}

  \begin{tikzpicture}[shorten >=1pt,
      node distance=2.0in,
      very thick]
    \tikzstyle{every node}=[base]
    \node (start)       []                       {};
    \node (domain) [xshift=-0.75in, yshift=0.75in, below right of=start] {
      \makecell[c]{
        \textbf{\large Domain Registration}\\
        Domain Features\\
        \Fref{sec:registration}
    }};
    \node (certificate) [right of=domain] {
      \makecell[c]{
        \textbf{\large Certificate Issuance}\\
        Certificate Features\\
        \Fref{sec:tls}
    }};
    \node (hosting) [right of=certificate] {
      \makecell[c]{
        \textbf{\large Website Deployment}\\
        Hosting Features\\
        \Fref{sec:hosting}
    }};
    \node (content) [right of=hosting] {
      \makecell[c]{
        \textbf{\large Content Publication}\\
        Natural Language and\\
        Perceptual Features \\
    }};
    \node (distribution) [right of=content] {
      \makecell[c]{
        \textbf{\large Content Distribution}\\
        Social Media Activity and \\ Consumption Features \\
    }};
    \node (end) [yshift=-0.75in, above right of=distribution] {};

    \draw[->] (start) -- node[above=.15in] {\large\emph{Time}} (end);
    \draw[-] (domain) |- (end);
    \draw[-] (certificate) |- (end);
    \draw[-] (content) |- (end);
    \draw[-] (hosting) |- (end);
    \draw[-] (distribution) |- (end);

    \node (below-domain)    [below of=domain, yshift=1.5in] {};
    \node (below-hosting)   [below of=hosting, yshift=1.5in] {};
    \node (below-end)       [below of=end, yshift=0.845in] {};
    \node (below-content)   [below of=content, yshift=1.25in] {};
    \node (below-distribution)   [below of=distribution, yshift=1.25in] {};
    \node (below-end-prior) [below of=end, yshift=.585in] {};

    \draw[|->,dashed,gray] (node cs:name=below-domain,anchor=south) -- node[below=.1in,pos=.275] {\large Our Work} (node cs:name=below-end,anchor=south);
    \draw[|->,dashed,gray] (node cs:name=below-content,anchor=south) -- node[below=.1in] {\large Prior Work} (node cs:name=below-end-prior,anchor=south);

  \end{tikzpicture}
} \caption{The
    lifecycle of a disinformation website, from domain registration to content
    distribution. We focus on features that are available as soon as the domain
    is registered, with progressively improving automated accuracy as the
    website deploys.} \label{fig:lifecycle} \end{figure*}

\section{Definitions and Scope} \label{sec:class_definitions} Definitions of
disinformation vary in academic literature and public
discourse~\cite{arXiv:1804.08559,tandoc2018defining}. Common components
include intent to deceive about facts~\cite{jack2017lexicon,
wardle2017information, fallis14functional, hernon95disinformation}, intent to
harm~\cite{wardle2017information}, and intent to prompt
distribution~\cite{lazer18science}. 

\subsection{Websites as Granularity of Study} We study disinformation at the
granularity of website domains, rather than individual articles, claims,
advertisements, social media accounts, or social media actions (e.g., posts
or shares). Websites are often used as distribution channels for
disinformation on social media platforms~\cite{guess2019fake2018,
guess2018selective, allcott2019trends, allcott2017social}, and there are several benefits to studying disinformation at the website level:

\begin{itemize} \item {\bf Early Warning.} It is possible, in principle, to
            identify disinformation websites before they begin to publish or
            distribute content. For example, an automated detection
            system might spot a new domain registration that looks like a
            local newspaper name, but has infrastructure overseas. A human
            moderator might then investigate and find there is no local
            newspaper with that name. Analysis of article content or social
            media activity, by contrast, can only occur much later in the
            disinformation lifecycle (\Fref{fig:lifecycle}).
    \item {\bf Longer-Term Value.} Disinformation articles and social media
        posts have an inherently limited lifespan, owing to the rapid news
        cycle~\cite{zubiaga2016analysing}. Disinformation websites, by
        contrast, can last for years.
    \item {\bf Platform Independence.} Identifying disinformation websites is
        feasible without access to a major online platform's internal account
        or activity data.
    \item {\bf Ecosystem Value.} A real-time feed of disinformation websites
        has value throughout the Internet ecosystem, similar to existing feeds
        of malware, phishing, and scam domains~\cite{safebrowsing}. Website
        data is immediately actionable for a diverse range of Internet
        stakeholders. Further, because websites are often components of
        multimodal disinformation campaigns, detection at the domain level can
        provide an investigative thread to untangle the rest of a
        disinformation campaign, including associated social media accounts
        and activities.  \end{itemize}

There are drawbacks and limitations associated with focusing on domains. Some
websites feature a mix of authentic and false news, complicating our class
definitions (see Section~\ref{subsec:class_definitions}). We also recognize
that websites are just one source of disinformation, and that private
communications and native social media content also play substantial roles in
exposing users to disinformation and instilling false beliefs. 

\subsection{Class Definitions} \label{subsec:class_definitions}
\paragraph{Disinformation} We define disinformation websites as websites that
appear to be news outlets with content about politics and current events, but
that operate in a manner significantly inconsistent with the norms, standards,
and ethics of professional journalism. Satire websites fall within
our definition of disinformation when the satire is not readily apparent to
users. This is an intentional definitional decision, since satire websites can
(and often do) mislead users~\cite{evon2019snopes, kasprak2019snopes}, and
since disinformation websites are known to sometimes rely on implausible
small-print disclaimers that they are satire~\cite{palma2017snopes,
palma2019snopes}. Our goal is to identify websites where users might benefit
from additional context or other interventions. 
We decline to use the term ``fake news,'' even though it may be a 
more apt description of the category of website that we study, because of the 
term’s political connotations and because the utility of our features is 
generalizable.

We focus on websites related to current events and politics because a significant proportion of the U.S. population has encountered these types of disinformation websites at least once~\cite{allcott2019trends,  guess2018selective, allcott2017social}, and certain groups of users (such as those over the age of 65) encounter them at high rates~\cite{guess19less}. 

\paragraph{Authentic News} We define authentic news websites, including those
with a partisan bias, as news websites that adhere to journalistic norms such
as attributing authors, maintaining a corrections policy, and avoiding
egregious sensationalism.

\paragraph{Non-news} We define a third category of non-news websites, which
are websites that primarily serve content other than news and do not represent
(or claim to represent) news outlets.

\section{Website Dataset} \label{sec:training} We used both current and
historical data to construct our dataset. We identified three classes, rather
than just two classes of fake and authentic news, both to facilitate feature
engineering and because we found that cleaner class separation improved
classification performance. We balanced the classes, 
including about 550 websites for each class.
\footnote{Our source code and dataset are available at https://github.com/ahounsel/disinfo-infra-public.}

We first constructed the
disinformation class, then constructed the other two sets with equal sizes for
balanced training and testing. Class sizes changed slightly over the course of
dataset construction, so the final datasets contain 551 disinformation sites, 
553 news sites, and 555 non-news sites. We recognize that the non-news website 
class would predominate in a real-time feed of domain, certificate, or social 
media events. Our rationale is that without balancing the dataset, the models 
we develop would minimize error by simply labeling every website as other.


\subsection{Disinformation Websites} We began by combining multiple
preexisting datasets of disinformation websites that had been manually labeled
by experts or published by news outlets, research groups, and professional
fact-checking organizations.  Specifically, we integrated the corpora from
CBS~\cite{cbs2017dont}, FactCheck.org~\cite{factcheck2018misinformation},
Snopes~\cite{lacapria2016snopes}, Wikipedia~\cite{wikipedia2019list},
\mbox{PolitiFact}~\cite{gillin2017politifacts}, and
\mbox{BuzzFeed}~\cite{silverman2016here,silverman2017these,silverman2017spite}.  We
also included websites that have been labeled as ``disinformation'' by
OpenSources, a collaborative academic project that manually assigned
credibility labels to news-like websites~\cite{zimdars2017opensources}.
Finally, we integrated the list of
disinformation websites compiled by Allcott et al. for their study on the
diffusion of disinformation on Facebook between 2015 and
2018~\cite{allcott2019trends}, which they also compiled from lists by
fact-checking organizations and academic sources.

We then manually filtered the list of websites, leaving only the websites that
satisfied our definition of disinformation
(Section~\ref{subsec:class_definitions}). Our
final dataset contains 758 disinformation websites. 575 (76\%) of the websites
are currently inactive: either unavailable, replaced with a parking page, or
repurposed for other kinds of abuse (e.g., spam or malware distribution). This
highlights the rapid turnover of disinformation websites in comparison to
authentic news websites.  Fortunately, we were able to reconstruct domain,
certificate, and hosting features for 368 (64\%) of these inactive websites
through Internet Archive Wayback Machine snapshots, the DomainTools API, and
the \texttt{crt.sh} Certificate Transparency log
database~\cite{internet-archive, domaintools, crtsh}. This resulted in a 
set of 551 disinformation websites.

\subsection{News Websites} We built a corpus of 553 authentic news websites,
randomly sampling 275 from Amazon's Alexa Web Information Service
(AWIS)~\cite{awis} and 278 from a directory of websites for local newspapers,
TV stations, and magazines~\cite{yin2018local}. From AWIS, we sampled websites
categorized as ``news'', excluding the 100 most popular websites out of
recognition that these websites likely have some distinct properties compared
to the long tail of news websites (e.g., high-quality and customized
infrastructure). From the local news dataset, we manually filtered to omit
websites that did not prominently display news (e.g., TV station websites that
served as channel guides).

\subsection{Other Websites} We built a set of 555 other websites by sampling
from Twitter's Streaming API~\cite{twitter-streaming-api}. We filtered for
tweets that contained a URL, extracted the domain name, and then used the
Webshrinker classification service~\cite{webshrinker} to assign labels based
on the Interactive Advertising Bureau's standardized website
categories~\cite{iab-taxonomy}. We excluded websites that belonged to the
``News'' and ``Politics'' categories.

%
%
%

\begin{table*}[ht!]
\centering
\renewcommand{\arraystretch}{1.1}
\small
\rowcolors{1}{}{lightgray}
    \resizebox{\textwidth}{!}{%
\begin{tabularx}{7in}{llXrl}
  \textbf{Name} & \textbf{Category} & \textbf{Description} & \textbf{Rank} & \textbf{Data Type} \\
  \hline
  News Keyword(s) in Domain & Domain & The domain name contains one or more keywords that imply it serves news (e.g., ``herald,'' ``tribune,'' or ``chronicle''). & 1 & Boolean\\
  Domain Name Length & Domain & The number of characters in the domain name. & 3 & Numeric\\
  ``News'' in Domain & Domain & The domain name contains the specific keyword ``news.'' & 8 & Boolean\\
  WHOIS Privacy & Domain & The domain registrant is using a WHOIS proxy service or registrar privacy option. & 9 & Boolean\\
  Registrar Name & Domain & The organization with whom the domain was registered. & 11 & Categorical \\
  Nameserver SLD & Domain & The second-level domain of the nameserver. & 14 & Categorical\\
  Nameserver AS & Domain & The autonomous system of the nameserver's IP address. & 16 & Categorical\\
  Registrant Organization & Domain & The organization of the registrant. & 17 & Categorical\\
  Registrant Country & Domain & The country of the registrant. & 19 & Categorical\\
  Time Since Domain Registration & Domain & The time elapsed since the domain was originally registered. & 21 & Numeric\\
  Domain Lifespan & Domain & The time period between the domain's initial registration and expiration dates. & 22 & Numeric\\
  Time to Domain Expiration & Domain & The time until the domain's registration expires. & 23 & Numeric\\
  Time Since Domain Update & Domain & The time since the domain's configuration was updated. & 25 & Numeric\\
  Nameserver Country & Domain & The country where the nameserver is located, using IP geolocation. & 27 & Categorical\\
  Novelty TLD & Domain & The TLD is novelty (e.g., \texttt{.news}, \texttt{.xyz}, or \texttt{.club}). & 29 & Boolean\\
  Digit in Domain & Domain & The domain name contains numeric characters. & 30 & Boolean\\
  Hyphen in Domain & Domain & The domain name contains a hyphen. & 31 & Boolean\\
  Domain Resolves & Domain & The domain name resolves to an IP address. & 32 & Boolean\\
  \arrayrulecolor{gray}\hline
  SAN Count & Certificate & The number of domains in the Subject Alternate Name extension field. & 2 & Numeric\\
  SAN Contains Wildcard & Certificate & The Subject Alternate Name extension field contains a wildcard entry for a domain. & 7 & Boolean\\
  Expired Certificate & Certificate & The certificate is expired. & 10 & Boolean\\
  Certificate Available & Certificate & A certificate is configured at the domain (i.e., a certificate is provided during a TLS handshake on the HTTPS port). & 12 & Boolean\\
  Self-signed Certificate & Certificate & The certificate is signed by the domain owner, not a CA. & 13 & Boolean\\
  Domain-validated Certificate & Certificate & The domain owner has obtained a certificate from a CA through domain validation rather than organization validation or extended validation. & 18 & Boolean\\
  Certificate Issuer Name & Certificate & The organization or individual who issued the certificate. & 24 & Categorical\\
  Certificate Issuer Country & Certificate & The country where the certificate was issued. & 26 & Categorical\\
  Certificate Lifetime & Certificate & The certificate's period of validity. & 28 & Numeric\\
  \arrayrulecolor{gray}\hline
  WordPress Plugins & Hosting & WordPress plugins used by the website. & 4 & Categorical\\
  Website AS & Hosting & The autonomous system of the website's IP address. & 5 & Categorical\\
  WordPress CMS & Hosting & The website uses WordPress as its content management system. & 6 & Boolean\\
  WordPress Theme & Hosting & The WordPress theme used by the website. & 15 & Categorical\\
  Website Country & Hosting  & The country where the website is located, using IP geolocation. & 20 & Categorical\\
  Website Available & Hosting & A website is hosted at the domain (i.e., content is returned in response to an HTTP request for the base URL, following redirects). & 33 & Boolean\\
  \arrayrulecolor{black}\hline
\end{tabularx}}
\caption{Domain, certificate, and hosting features that our model uses to classify a website as authentic news, disinformation, or other. Features are ranked by Gini importance in our random forest model.}
\label{tab:features}
\end{table*}

\section{Feature Engineering} \label{sec:features} We engineered 33 features
to distinguish disinformation, authentic news, and other websites (see
Table~\ref{tab:features}). In this section, we describe exemplary features.

\subsection{Domain Features} \label{sec:registration} Eighteen features
related to a website's domain name, registration, or DNS configuration. Domain
names were valuable for distinguishing authentic and disinformation websites
from other websites; these classes generally used a domain name with a
news-related keyword like ``news'', ``herald'', or ``chronicle''. Domain
registrars showed distinct usage patterns between authentic and disinformation
websites: two low-cost and consumer-oriented registrars (Namecheap and Enom)
were much more common among disinformation websites, while business-oriented registrars like Network Solutions, MarkMonitor, and CSC were more common for authentic news websites. We also found that authentic news websites were much more likely to have old domain registrations with far-off expiration dates, and that disinformation websites were much more likely to use new TLDs like \texttt{.news}. We observe that disinformation creators trying to evade detection based on domain features would need to plan in advance, establish business relationships, and bear other costs that may be deterring.

\begin{figure*}[ht!] \centering \begin{subfigure}[b]{0.45\textwidth}
\centering \includegraphics[width=0.9\linewidth]{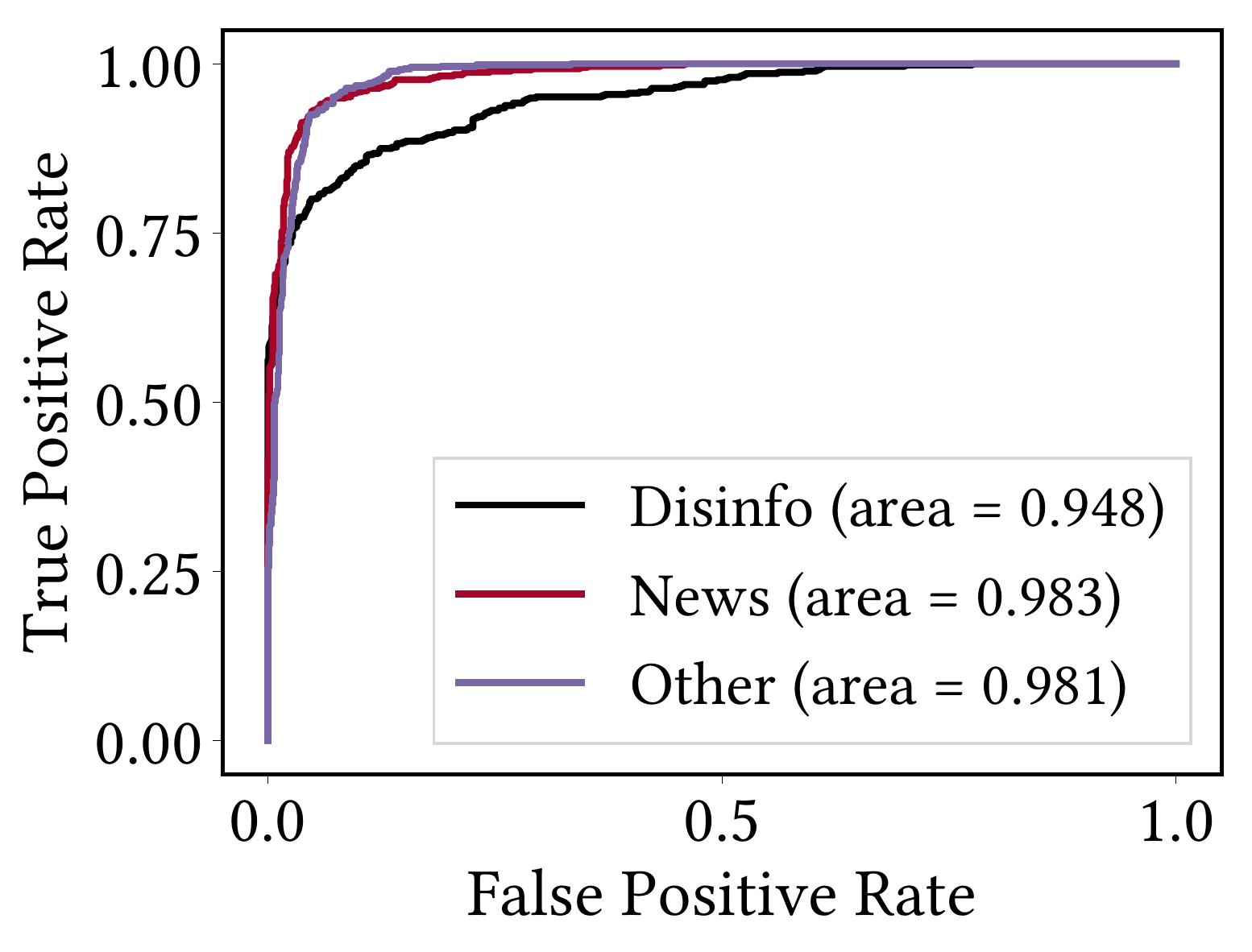}
\caption{ROC Curves \label{fig:classifier_class_roc}} \end{subfigure} \hfill
\begin{subfigure}[b]{0.45\textwidth} \centering
\includegraphics[width=0.9\linewidth]{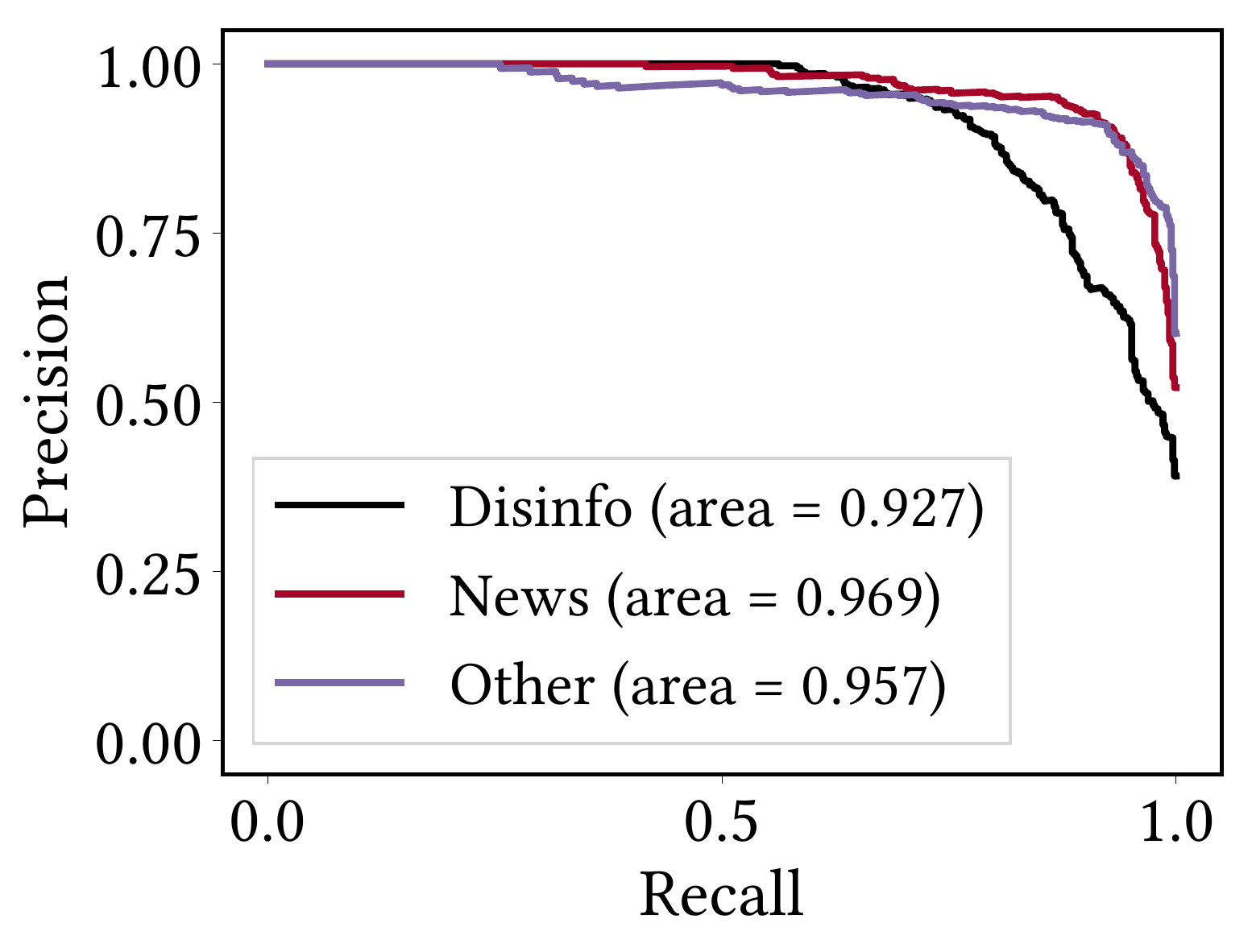} \caption{P-R
Curves \label{fig:classifier_class_pr}} \end{subfigure} \\
\begin{subfigure}[b]{0.45\textwidth} \centering
    \includegraphics[width=0.9\linewidth]{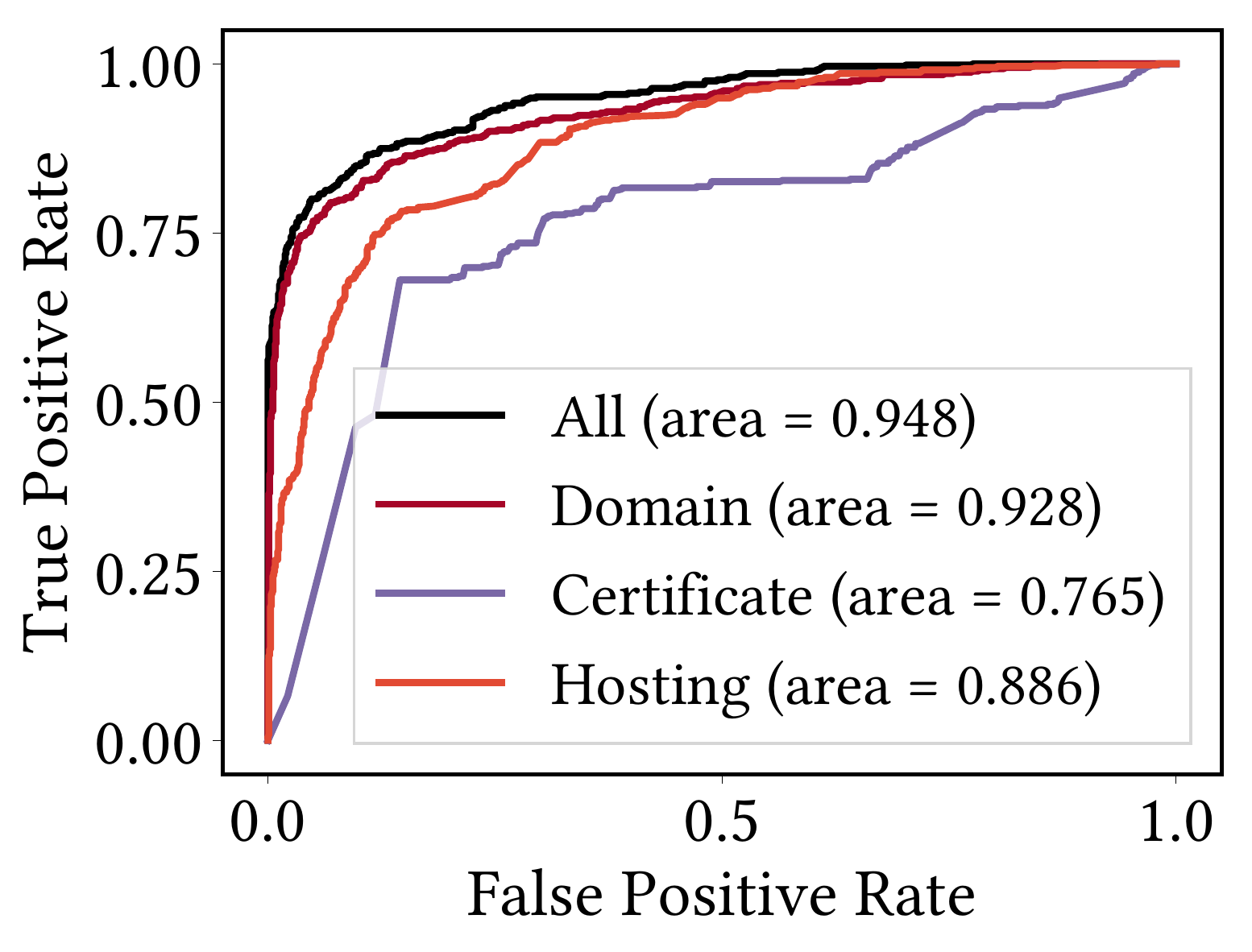}
\caption{ROC Curves \label{fig:classifier_timeline_roc}} \end{subfigure}
\hfill \begin{subfigure}[b]{0.45\textwidth} \centering
    \includegraphics[width=0.9\linewidth]{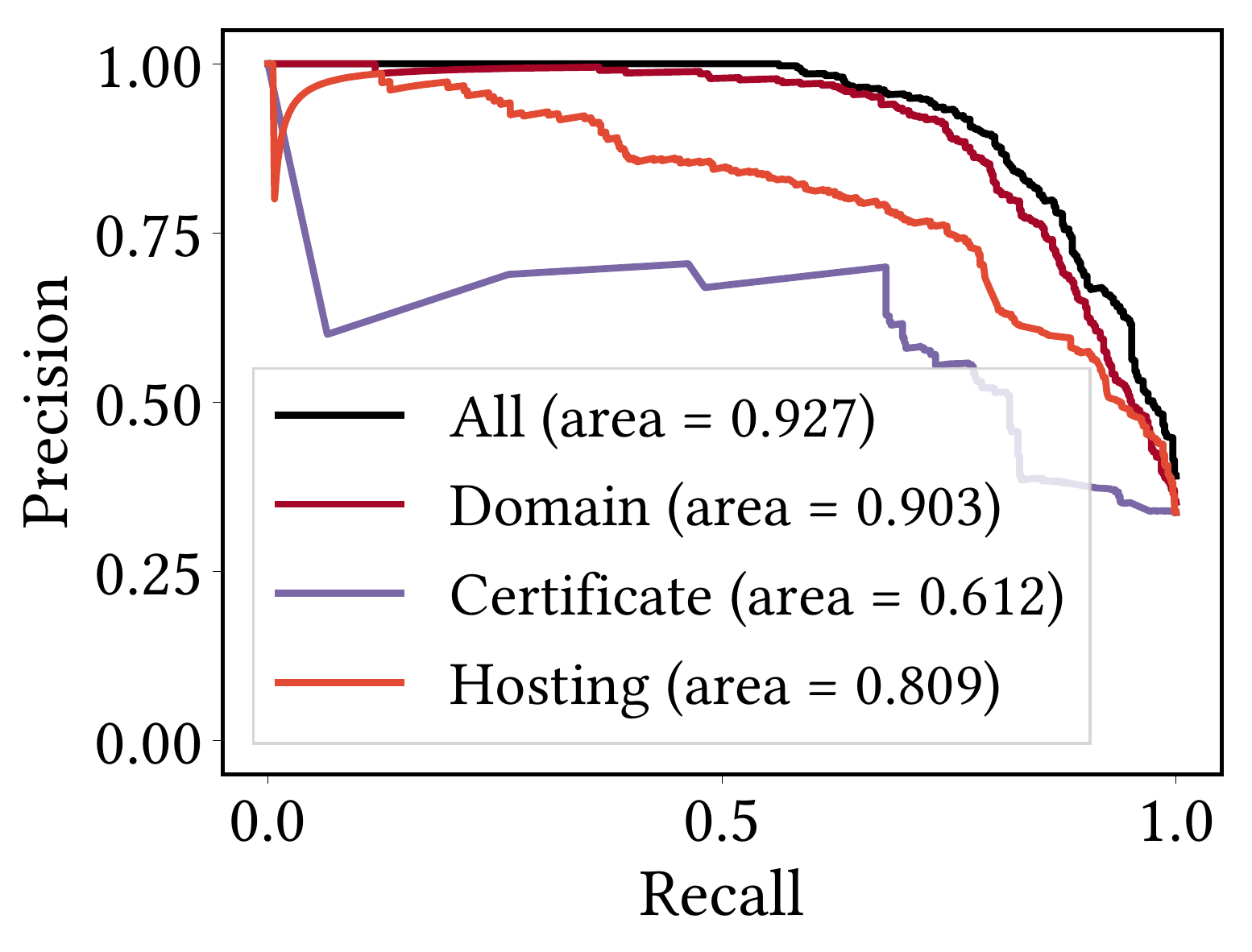}
\caption{P-R Curves \label{fig:classifier_timeline_pr}} \end{subfigure}
\caption{Model performance on the curated dataset. Figures
\ref{fig:classifier_class_roc} and \ref{fig:classifier_class_pr} show
performance on all three classes using all feature types. Figures
\ref{fig:classifier_timeline_roc} and \ref{fig:classifier_timeline_pr} show
performance on the disinformation class using subsets of features.}
\label{fig:classifier_class} \end{figure*}

\subsection{Certificate Features} \label{sec:tls} Nine features related to a
website's TLS/SSL certificate. One key feature was the number of domains that
a certificate covers (based on the Subject Alternative Name field).  We found
that news websites often had more domains in their certificates than
disinformation websites, because parent news organizations used one
certificate to cover their subsidiaries. We also found, though, that some
disinformation websites had a large count of domains in their certificates
attributable to low-cost hosting providers that deployed shared certificates.

\subsection{Hosting Features} \label{sec:hosting} Six features pertained to a
website's hosting infrastructure. Using BGP routing tables, we identified the
autonomous system hosting each website's IP address and found that mass-market
hosting providers like GoDaddy and Namecheap were much more common among
disinformation websites, while premium, business-oriented hosting providers
like Incapsula were more common among authentic news websites. We also
geolocated IP addresses using the MaxMind GeoLite2 database and found that,
contrary to our expectations, geolocation was not a valuable feature because most
websites in all three classes were U.S.-based.

The presence of web trackers and low-quality advertisements may also be useful features, although we did not find that a feature based on websites' Google Analytics IDs improved model accuracy. We leave the evaluation of other trackers to future work.

\section{Model} \label{subsec:model} We selected a multi-class random forest
model because we expected that feature interactions would contribute to
performance and that model interpretability would be important for evaluation
and plausible deployment. We conducted a randomized hyperparameter search over
a wide range of values, then selected values that achieved the best average
accuracy based on 250 iterations with five-fold cross-validation.

Our model was able to distinguish the three classes and classify websites
accurately; the mean ROC AUCs for authentic news websites, disinformation
websites, and other websites were 0.98, 0.95, and 0.98 respectively, and the
mean precision-recall AUCs were 0.97, 0.93, and 0.96 respectively. The
performance of our model surpasses the prior work~\cite{baly2018predicting}
and is comparable to concurrent work~\cite{chen2020}. We present ROC and
precision-recall figures in Figure~\ref{fig:classifier_class}.

We evaluated the importance of the domain, certificate, and hosting feature
categories by training and testing a standalone model for each category.
Figure~\ref{fig:classifier_class} presents the comparative performance of
these models. We found that domain features predominantly drove classification
performance. This result is promising, because domain features are available
very early in a disinformation website's lifecycle; domain registrars (among
other Internet stakeholders) could intervene or begin heightened monitoring
when a suspicious domain appears.  We found that hosting features accomplished
moderate performance, while certificate features contributed little to
classification performance.

\begin{figure}[t!] \centering
    \includegraphics[width=0.8\linewidth]{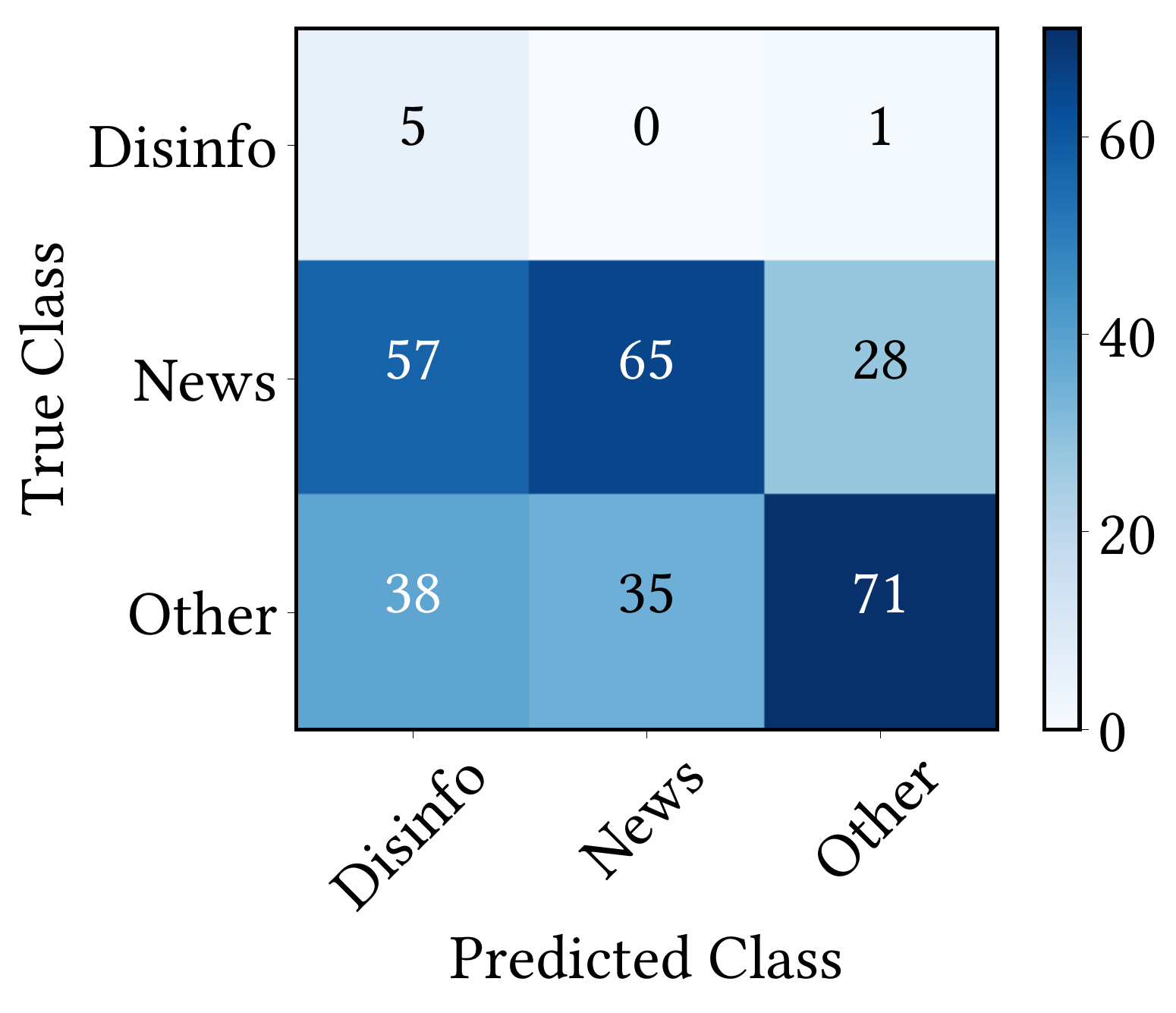}
    \caption{Performance on 100 manually labeled websites from each class,
    sampled from real-time pilot data.} \label{fig:live_confusion_matrix}
\end{figure}

\section{Pilot Real-Time Deployment}
We conducted a pilot real-time deployment of our classifier to understand how 
well its performance generalizes. Our implementation used a commodity server 
to ingest new domains from DomainTools~\cite{domaintools} (which tracks domain 
registrations), CertStream~\cite{certstream} (which tracks new TLS/SSL 
certificates), the Twitter Streaming API~\cite{twitter-streaming-api}, and the 
Reddit API~\cite{reddit-api}, then another commodity server to collect 
infrastructure data, generate features, and output a classification. We ran the 
system for 5 days, classifying 1,326,151 websites. To simulate our 
classifier on a platform that has access to all features, we 
randomly sampled 300 websites that appeared on Twitter---100 from each detected 
class---and manually labeled to evaluate performance. We present a confusion 
matrix in Figure~\ref{fig:live_confusion_matrix}.

Our model's precision on the disinformation class (0.05) was sufficient for
plausible deployment. We were able to rapidly discard false positives and,
just in our small-scale pilot deployment, discovered two disinformation
websites that had not been previously reported in any public venue.

We cautiously note that model performance radically degraded in comparison to
our prior evaluation, similar to results in concurrent
work~\cite{chen2020}. We attribute this difference to three potential causes.
There is a massive class imbalance inherent in real-world classification---the
overwhelming majority of websites do not relate to news, and most news links
shared on social media are for authentic news websites rather than disinformation websites.
As a result, even with good performance, false positives may dominate
true positives. Also, as training data gets older, the features for current
websites may change. This may create a gap between the features observed in
the training data and current disinformation websites.

It is also possible that our model is picking up on artifacts in our training
dataset to classify websites. For example, $\sim$34\% of the disinformation
websites in our training data are active, whereas all of the news websites are
active. If there are differences in the features of inactive websites that we
reconstructed features for and active websites, then our classifier may be
using the wrong signal to distinguish disinformation websites from news
websites. We leave an evaluation of our classifier that is only trained
on active websites to future work.

\section{Evasion Resistance} \label{sec:discussion} Disinformation website operators
will be motivated to evade detection.  In other areas of online abuse, such as
spam, phishing, and malware, online platforms are constantly developing new defensive measures to keep up with advances in adversary capabilities. We expect
that disinformation will follow a similar cat-and-mouse pattern of defense and
evasion.

Our model uses features that provide a degree of asymmetric advantage in
identifying disinformation websites, since a website that seeks to evade
detection must make changes to its infrastructure.  Some features will be
relatively easy to evade; for example, a website can easily change a WordPress
theme or renew an expired TLS certificate. Fortunately, many of the most
important features that our model relies on are difficult or costly to evade.

As an example, consider one of the most predictive features: the lifespan of a
website's domain. Evading that feature requires either significant advance
planning or purchasing an established domain.
Evading certain other features incurs monetary costs, like purchasing a
certificate from a reputable issuer, registering a domain for a longer time,
switching to a more expensive non-novelty TLD, or migrating to a more
trustworthy hosting provider. Evading other features incurs technical costs:
obtaining and installing a correctly configured, reputably issued TLS
certificate, for instance, imposes some operational cost.  Finally, evading many of our
model's features might reduce the effectiveness of the disinformation
campaign.  For example, a top ranked feature is whether a domain contains news
keywords.  Removing those keywords from the domain name could diminish the
credibility of the website and lead to less exposure on social media.

\section{Related Work} \label{sec:related-work} Our approach is inspired by
the information security literature on detecting malicious websites. Prior
work has demonstrated the value of infrastructure features for identifying
spammers~\cite{ramachandran2006understanding,hao2009detecting},
botnets~\cite{gu2008botminer,ramachandran2006revealing}, and
scams~\cite{konte2009dynamics,hao2010internet,hao2016predator}.

Efforts at classifying disinformation have predominantly used natural
language features
(e.g.,~\cite{qazvinian2011rumor,arXiv:1708.07104,horne2019this,shu2018fake,obrien2018the,wu2010distortion,chen2015towards,conroy2016automatic,
rubin2014truth,
rubin2015deception,zhang2012an,markowitz2014linguistic,potthast2018stylometric,afroz2012detecting,kumar2016disinformation,mihalcea2009lie,rashkin2017truth}).
Several projects have relied on social graph features (e.g.,
~\cite{gupta2013faking,zhao2015enquiring,yang2012automatic,ma2017detect,wu2015false}).

Baly et al. predicted news website
factuality by examining the domain name, article text, associated Wikipedia
and Twitter pages, and web traffic statistics for a combined accuracy of about
0.5~\cite{baly2018predicting}. Chen and Freire
developed a system for discovering political disinformation websites on
Twitter that uses article text and markup~\cite{chen2020}. The authors report
a mean ROC AUC of 0.97 on historical data and significantly lower performance
in a trial real-time deployment.

In comparison to the disinformation detection literature, we contribute a new
set of infrastructure features that do not rely on content or distribution. We
demonstrate website classification with state-of-the-art performance and
characterize how performance degrades in a real-world deployment setting.

\section{Conclusion} Our work demonstrates a promising new set of features for
detecting disinformation websites and a real-time classification system could
feasibly be deployed by online platforms. Future work on disinformation
detection should examine how infrastructure features interact with natural
language and social sharing features. Our work also highlights the significant
and unexplored practical challenges of real-world deployment for
disinformation detection systems. As a research community, we have an
opportunity to support the free and open speech environment online by enabling
responses to disinformation. But we have to move beyond detection methods that
are only strong on paper and begin addressing the hard problems associated
with real-world deployment.

\section*{Acknowledgments}
We thank our paper shepherd Jed Crandall and our anonymous reviewers for their 
feedback. We also thank DomainTools for providing access to their services for 
our research. This work was funded in part by National Science Foundation Award 
TWC-1953513.

\renewcommand*{\bibfont}{\small}

\clearpage
\balance
\printbibliography

\clearpage
\sloppy
\appendix
\section{Appendix}

\subsection{Domain Proxy Keywords}
We first computed the most popular values of the ``registrant organization'' field from WHOIS records in our training data.
Then, we manually extracted keywords from known domain proxy services that were highly ranked.
We used the resulting list of keywords as a heuristic to determine if new domains use WHOIS privacy services:

\begin{table}[h!]
  \centering
  \renewcommand\arraystretch{0.95}
  \small
  \begin{tabularx}{0.95\columnwidth}{XXX}
    domain         & protect  & whois \\
    guard          & proxy    &       \\
    privacy        & redacted &
  \end{tabularx}
\end{table}

\subsection{News-Indicating Keywords}
We used the following 163 keywords with the DomainTools Brand Monitor API to retrieve newly registered domains that may be news websites.
We derived the keywords by manually examining our training datasets of authentic and fake news websites for common terms related to news, media, information, or publishing.

\begin{table}[ht!]
  \centering
  \renewcommand\arraystretch{0.95}
  \small
  \begin{tabularx}{\linewidth}{XXXX}
  	24 & 365 & abc & action \\
  	activist & advance & alert & alliance \\
  	alternative & america & associate & blast \\
  	blog & box & breaking & brief \\
  	bulletin & business & buzz & byte \\
  	caller & cbs & channel & christian \\
  	chronicle & citizen & city & club \\
  	cnn & conservative & corner & county \\
  	courier & currant & daily & democracy \\
  	dig & dispatch & division & edition \\
  	editor & election & empire & epoch \\
  	evening & examiner & express & extra \\
  	fact & feed & file & finesser \\
  	flash & focus & fox & free \\
  	fresh & gazette & global & guardian \\
  	hangout & headline & herald & hq \\
  	hub & idea & independent & index \\
  	info & inquire & insider & interesting \\
  	international & item & journal & leak \\
  	learn & ledger & liberal & liberty \\
  	live & local & mag & maga \\
  	mail & media & metro & movement \\
  	nation & nbc & network & now \\
  	observer & page & paper & patriot \\
  	pioneer & pipe & plug & politic \\
  	post & press & progress & proud \\
  	publish & radio & react & read \\
  	record & region & religion & report \\
  	republic & review & rumor & scoop \\
  	sentinel & share & sharing & show \\
  	spirit & spot & spotlight & standard \\
  	star & state & statesman & stories \\
  	story & studio & sun & surge \\
  	syndicate & telegram & telegraph & television \\
  	times & today & top & trend \\
  	tribune & truth & tv & twenty-four \\
  	twentyfour & uncut & underground & union \\
  	update & us & usa & view \\
  	viral & vision & washington & watch \\
  	weekly & wire & your & zine \\
  	zone & & &
  \end{tabularx}
\end{table}

\end{document}